\documentclass[%
 reprint,
superscriptaddress,
 amsmath,amssymb,
 aps,
]{revtex4-2}

\usepackage{graphicx}
\usepackage{dcolumn}
\usepackage{bm}
\usepackage{color}
\usepackage{xspace}
\usepackage{multirow}
\usepackage{hyperref}

\usepackage{here}

\definecolor{green}{rgb}{0,0.6,0.1}

\newcommand{\xx}{$d_{x^2-y^2}$\xspace}

\newcommand{\zz}{$d_{3z^2-r^2}$\xspace}
\newcommand{\QE}{{\textsc{Quantum ESPRESSO}}\xspace}

\newcommand{\Rb}{RbCa$_2$NiO$_3$\xspace}
\newcommand{\Br}{$A_{2}$NiO$_{2}$Br$_2$\xspace}

\newcommand{\cation}{Ba$_{0.5}$La$_{0.5}$\xspace}

\begin{document}

\preprint{APS/123-QED}

\title{
Magnetic exchange coupling in cuprate-analog $d^9$ nickelates
}

\author{Yusuke Nomura}
\email{yusuke.nomura@riken.jp}
\affiliation{ 
RIKEN Center for Emergent Matter Science, 2-1 Hirosawa, Wako, Saitama 351-0198, Japan 
}

\author{Takuya Nomoto}
\affiliation{
Department of Applied Physics, The University of Tokyo,7-3-1 Hongo, Bunkyo-ku, Tokyo 113-8656
}

\author{Motoaki Hirayama}
\affiliation{ 
RIKEN Center for Emergent Matter Science, 2-1 Hirosawa, Wako, Saitama 351-0198, Japan 
}

\author{Ryotaro Arita}
\affiliation{ 
RIKEN Center for Emergent Matter Science, 2-1 Hirosawa, Wako, Saitama 351-0198, Japan 
}
\affiliation{
Department of Applied Physics, The University of Tokyo,7-3-1 Hongo, Bunkyo-ku, Tokyo 113-8656
}

\date{\today}

\begin{abstract}
Motivated by the recent discovery of superconductivity in doped NdNiO$_2$, we study the magnetic exchange interaction $J$ in layered $d^9$ nickelates from first principles.
The mother compounds of the high-$T_{\rm c}$ cuprates belong to the charge-transfer regime in the Zaanen-Sawatzky-Allen diagram and have $J$ larger than 100 meV. While this feature makes the cuprates very different from other transition metal oxides, it is of great interest whether layered $d^9$ nickelates can also have such a large $J$. However, one complexity is that NdNiO$_2$ is not a Mott insulator due to carrier doping from the block layer. To compare the cuprates and $d^9$ nickelates on an equal basis, we study \Rb and \Br ($A$: a cation with the valence of $2.5+$), which were recently designed theoretically by block-layer engineering. These nickelates are free from the self-doping effect and belong to the Mott-Hubbard regime. We show that these nickelates share a common thread with the high-$T_{\rm c}$ cuprates in that they also have a significant exchange interaction $J$ as large as about 100 meV.
\end{abstract}

\maketitle


\section{Introduction}


The discovery of superconductivity in doped nickel oxides Nd$_{0.8}$Sr$_{0.2}$NiO$_2$~\cite{Li_2019,Sawatzky19} has attracted intensive interests 
both in experiment~\cite{Hepting_2020,Q_Li_2020,Zhou_2020,Fu_arXiv,Lee_2020,D_Li_2020,Zeng_arXiv,Goodge_arXiv,BX_Wang_2020,Q_Gu_arXiv,Osada_2020} and theory~\cite{Hepting_2020,Botana_2020,Sakakibara_2020,Hirsch_2019,
Nomura_2019,Hirayama_2020,Gao_arXiv,Singh_arXiv,
Jiang_2020,Ryee_2020,HuZhang_2020,GM_Zhang_2020,Z_Liu_2020,
Wu_2020,Been_arXiv,Lang_arXiv,Leonov_2020,Leonov_arXiv_2,
Werner_2019,Petocchi_arXiv,Y_Gu_2020,Liang_Si_2020,Lechermann_2020,Lechermann_arXiv,Karp_2020,Kitatani_arXiv,Y_Wang_arXiv,
Zhang_2020,LH_Hu_2019,Chang_arXiv,
Z_Wang_arXiv,
P_Jiang_2019,Choi_2020,
Geisler_2020,He_2020,Bernardini_2020c,
Talantsev_2020,T_Zhou_2020,Bernardini_2020,
Bernardini_2020b,Olevano_2020,Choi_arXiv,Adhikary_arXiv,Nica_arXiv}, 
because the nickelate might be an analog of the well known high-$T_{\rm c}$ superconductor, cuprates. 
Recently, the doping dependence has been explored both theoretically~\cite{Kitatani_arXiv} and experimentally~\cite{D_Li_2020,Zeng_arXiv},  
and the presence of the superconducting dome has been confirmed~\cite{D_Li_2020,Zeng_arXiv}.
The maximum superconducting transition temperature $T_{\rm c}$ is about 15 K, not very high compared to that of the high-$T_{\rm c}$ cuprates. 
However, because the Bardeen-Cooper-Schrieffer (BCS) phonon mechanism cannot explain the observed $T_{\rm c}$~\cite{Nomura_2019}, the superconducting mechanism is most likely unconventional, in which the electron correlations play an important role~\cite{Sakakibara_2020,Wu_2020,Kitatani_arXiv,Adhikary_arXiv}. 
A recent observation of $d$-wave like superconducting gap also supports this scenario~\cite{Q_Gu_arXiv}.
Here, a natural question arises: is there any possibility to realize $T_{\rm c}$ as high as the cuprates in nickelates?

In the cuprates, the superconductivity emerges by doping carriers into the antiferromagnetic Mott insulator having a large magnetic exchange coupling $J$ ($\sim$130 meV)~\cite{Lee_Nagaosa_Wen_2006}. 
One of the reasons for the large $J$ is because the cuprates belong to the charge-transfer type in the Zaanen-Sawatzky-Allen diagram~\cite{Zaanen_1985}, and the charge-transfer energy $\Delta_{dp}$ (the energy difference between the copper $3d$ and oxygen 2$p$ orbitals) is small among transition metal oxides.
Although the mechanism of the high-$T_{\rm c}$ superconductivity is highly controversial, the large $J$ is a plausible factor in enhancing the $d$-wave superconductivity in the cuprates~\cite{Ogata_2008}.
This large value of $J$ is certainly a characteristic feature of the cuprates, which makes the cuprates very different from other transition metal oxides.

On the other hand, in the case of the nickelate NdNiO$_2$, $\Delta_{dp}$ is larger than that of the cuprates~\cite{Lee_2004}. 
Thus, naively, we expect smaller $J$ for nickelates. 
Indeed, a recent experimental estimate using the Raman spectroscopy gives $J = 25$ meV~\cite{Fu_arXiv}.
However, it should be noted that the origin of small $J$ in NdNiO$_2$ may be ascribed to another notable difference from the cuprates, namely, NdNiO$_2$ is not a Mott-insulator due to the self-doping effect.
In NdNiO$_2$, orbitals in the Nd layer form extra Fermi pockets on top of the large Fermi surface formed by the Ni 3\xx orbital, and the Ni 3\xx orbital is hole-doped,
i.e., the filling of the Ni $3d$ orbitals deviates from $d^9$~\cite{Lee_2004,Botana_2020,Sakakibara_2020,Gao_arXiv,Liang_Si_2020,Olevano_2020}. 
The self-doping naturally explains the absence of Mott-insulating behavior in NdNiO$_2$. 
Although it has been shown that the Ni 3\xx orbital forms a two-dimensional strongly-correlated system~\cite{Nomura_2019}, $J$ at the $d^9$ configuration with half-filled \xx orbital is masked by the self-doping.
The experimental estimate should be understood as the $J$ value including the effect of the self-doping, not the $J$ value at the ideal $d^9$ configuration. 
One of the reasons for the controversy in theory about the size of $J$~\cite{Jiang_2020,Ryee_2020,HuZhang_2020,GM_Zhang_2020,Z_Liu_2020,Wu_2020,Been_arXiv,Lang_arXiv,Leonov_2020,Leonov_arXiv_2} is ascribed to the ambiguity in calculating $J$ (whether we calculate $J$ at $d^9$ filling or $J$ including the self-doping effect).  
In any case, it is a non-trivial problem whether we can justify the mapping onto a simple spin model to understand the property of NdNiO$_2$. 
This fact makes NdNiO$_2$ an imperfect analog of the cuprates.

Recently, there was a proposal to design cuprate-analog nickelates without the complication of the self-doping~\cite{Hirayama_2020} \footnote{See also Refs.~\cite{Bernardini_2020b,Nica_arXiv} for other attempts to find nickelate superconductors.}. 
Since NdNiO$_2$ is a layered material, one can systematically propose nickelate family materials by changing the composition of the ``block-layer" \cite{Tokura_1990} between NiO$_2$ layers.  
Proposed dynamically stable nickelates have smaller Fermi pockets of the block-layer orbitals than NdNiO$_2$. 
In some materials, the self-doping is completely suppressed, and the ideal $d^9$ system with half-filled 3\xx orbital is realized.
An {\it ab initio} estimate of Hubbard $U$ using the constrained random-phase approximation (cRPA)~\cite{Aryasetiawan_2004} shows that the correlation strength $U/t$ ($t$: nearest-neighbor hopping) is comparable to that of cuprates~\cite{Hirayama_2020}. 
Therefore, once such nickelates are synthesized, the mother compounds will be a Mott insulator similarly to the cuprates, and the effective model becomes the Heisenberg model, which gets rid of the ambiguity in calculating $J$.

In this paper, we study the strength of $J$ in the two ideal $d^9$ nickelates, which are free from the self-doping (see Sec~\ref{sec_materials} for the details of the materials). 
We estimate the $J$ value by the following three methods~\cite{Lichtenstein_2013}. 
First, we start from a single-orbital Hubbard model derived in Ref.~\onlinecite{Hirayama_2020} and then evaluate $J$ by the expansion in terms of $t/U$. Second, we perform an energy mapping between the classical Heisenberg model and the total energy of different magnetic configurations calculated by the LDA+$U$ (LDA: local density approximation) method. 
Third, we employ a scheme based on the so-called local force theorem. 
Hereafter, we simply call these three methods ``strong-coupling expansion", ``energy mapping method", and ``local force approach", respectively.
We show that the three independent estimates show reasonable agreement and conclude that the $d^9$ nickelates have sizeable $J$ (about 100 meV), 
which is not far smaller than that of the cuprates.
Therefore, the proposed $d^9$ nickelates provide an interesting playground to explore the cuprate-analog high-$T_{\rm c}$ superconductivity. 

The paper is organized as follows. 
In Sec.~\ref{sec_materials}, we introduce two ideal $d^9$ nickelates, \Rb and \Br ($A$: a cation with the valence of $2.5+$) and discuss the advantage over NdNiO$_2$. In Sec.~\ref{sec_methods}, we explain the three methods employed in the present study, and we show the results in Sec.~\ref{sec_results}. Section~\ref{sec_summary} is devoted to the summary.

\begin{figure*}[tb]
\vspace{0.0cm}
\begin{center}
\includegraphics[width=0.98\textwidth]{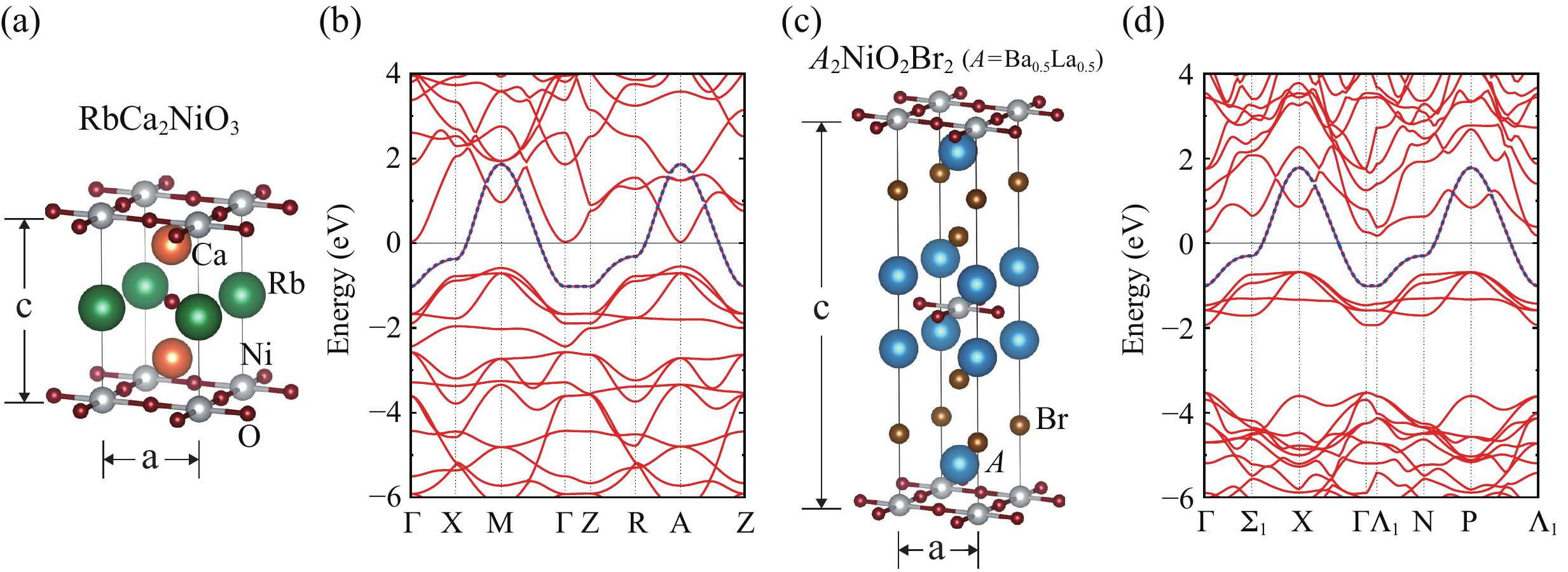}
\caption{
Crystal structure of (a) \Rb and (c) \Br ($A$: a cation with the valence of $2.5+$) and the paramagnetic DFT band structure [(b) \Rb and (d) \Br ($A$ = \cation)].
The blue dotted curves are the Wannier band dispersion of the Ni 3\xx single-orbital Hamiltonian. 
In (b) and (d), the consistent ${\bf k}$ path is employed: $(0,0,0)$ $\rightarrow$ $(\pi/a,0,0)$ $\rightarrow$ $(\pi/a,\pi/a,0)$ $\rightarrow$  $(0,0,0)$ $\rightarrow$ $(0,0,\pi/c)$ $\rightarrow$ $(\pi/a,0,\pi/c)$ $\rightarrow$ $(\pi/a,\pi/a,\pi/c)$ $\rightarrow$  $(0,0,\pi/c)$ (The symbols are different because the primitive cells of \Rb and \Br are tetragonal and bace-centered tetragonal, respectively).
}
\label{Fig_crystal_structure}
\end{center}
\end{figure*}

\section{Materials: $d^9$ nickelates} 
\label{sec_materials}

In Ref.~\cite{Hirayama_2020}, various layered nickelates have been systematically proposed. 
They are classified into ``1213", ``1214", ``H$_2$", and ``G" families, depending on the composition and the type of the block-layer~\cite{Tokura_1990}.
Among the four families, the compounds without the self-doping exist in the 1213 and G families.  
We here take \Rb and \Br ($A$: a cation with the valence of $2.5+$) for a representative of the ideal $d^9$ nickelates belonging to 1213 and G families, respectively (see Figs.~\ref{Fig_crystal_structure}(a) and (c) for the crystal structure). 
In the following, we employ \cation as $A$. 
The phonon calculations have shown that both \Rb and \Br ($A$ = \cation) are dynamically stable~\cite{Hirayama_2020}. 
We take the crystal structure optimized in Ref.~\cite{Hirayama_2020}, and perform density-functional theory (DFT) calculations~\footnote{Here, we ignore the interface effect~\cite{Geisler_2020,He_2020,Bernardini_2020c} and consider the bulk property.
Note that the thickness of the film reaches around 10 nm and there are several tens of NiO$_2$ layers in the sample~\cite{Lee_2020}.}. 
Figs.~\ref{Fig_crystal_structure}(b) and \ref{Fig_crystal_structure}(d) show the paramagnetic DFT band structure for \Rb and \Br ($A$ = \cation), respectively.
As is shown in Ref.~\cite{Hirayama_2020}, only the Ni 3\xx orbital crosses the Fermi level. 
As far as the topology of the band structure is concerned, these systems are more similar to the cuprates than NdNiO$_2$. 

The advantages of studying these nickelates rather than NdNiO$_2$ are as follows.
First, it is still controversial whether the role of Nd-layer (block-layer) orbitals is essential or not. 
If the hybridization between Ni $3d$ and Nd-layer orbitals is substantial, the Nd-layer orbitals are not only a charge reservoir, but they might give Kondo-like physics~\cite{Sawatzky19,Hepting_2020,GM_Zhang_2020,Z_Wang_arXiv,Y_Gu_2020}.  
In the cases of the $d^9$ nickelates, \Rb and \Br ($A$ = \cation), the block-layer orbitals do not show up at the Fermi level, and this controversy can be avoided. 
We can also exclude the possible role of the $4f$ orbitals with localized moments proposed in Refs.~\cite{P_Jiang_2019,Choi_2020}.

Another controversial issue for NdNiO$_2$ is to which orbitals the doped holes go ($d^9 \underline{L}$ vs. $d^8$, where $\underline{L}$ denotes a hole in a ligand oxygen).
In the case of the cuprates (charge-transfer insulator), the holes are doped into the oxygen $2p$ orbitals. 
On the other hand, the nickelates have larger $\Delta_{dp}$ and are classified as Mott-Hubbard type~\cite{Hepting_2020,Jiang_2020,Nomura_2019,HuZhang_2020,Fu_arXiv,Goodge_arXiv}.
Because there is nonzero hybridization between Ni 3\xx and O $2p$ orbitals, some of the holes should be doped into oxygen $2p$ orbitals~\cite{Hirsch_2019,Karp_2020,Lang_arXiv}. However, the amount should be smaller than that of the cuprates.

When the system is Mott-Hubbard type and the holes mainly reside in the Ni $3d$ orbitals, another issue arises: which model is more appropriate, the single-orbital or multi-orbital model? 
In other words, whether the doped $d^8$ configuration favors high-spin state or low-spin state.  
If the crystal field splitting between Ni 3\xx and the other $3d$ orbitals is much larger than the Hund's coupling, holes stay within the Ni 3\xx orbital, and the single-orbital model is justified. 
On this issue, several studies insist that Ni $3d$ multi-orbital nature cannot be 
ignored~\cite{Jiang_2020,Zhang_2020,Werner_2019,Petocchi_arXiv,LH_Hu_2019,Lechermann_2020,Lechermann_arXiv,Y_Wang_arXiv,Chang_arXiv,Choi_arXiv}.
To resolve this issue, we certainly need more experimental evidences. 
In the cases of \Rb and \Br ($A$ = \cation), compared to NdNiO$_2$, the Ni 3\xx orbital is more isolated in energy space from the other $3d$ orbitals [see Figs.~\ref{Fig_crystal_structure}(b) and \ref{Fig_crystal_structure}(d)]: 
In the case of NdNiO$_2$, due to the dispersion along the $k_z$ direction, the position of the Ni \zz band becomes close to the Fermi level on the $k_z = \pi/c$ plane; however, such $k_z$ dependence is much weaker in \Rb and \Br ($A$ = \cation).
Considering also the above-mentioned absence of the complication from the self-doping, in this study, we adopt the single-orbital Hubbard model as a minimal model for \Rb and \Br ($A$ = \cation). 
In the absence of the carrier doping, we can further map onto the spin model with the exchange coupling $J$.

\section{Methods}
\label{sec_methods}

Here, we introduce three different methods to estimate $J$
(see eg., Ref.~\cite{Lichtenstein_2013} for the ideas behind the three methods). 
We employ the following convention for the spin Hamiltonian: ${\mathcal H} = \sum_{\langle i,j \rangle } J_{ij} {\bf S}_i \cdot {\bf S}_j$, where $\langle i,j \rangle$ is the bond consisting of sites $i$ and $j$, and ${\bf S}_i$ is the spin-1/2 operator at site $i$. 
$J$ stands for the nearest-neighbor $J_{ij}$ interaction in the NiO$_2$ layer.

\subsection{Strong-coupling expansion} 
\label{sec_method_superexchange}


When the single-orbital Hubbard model is a good description, 
the magnetic interactions in the Mott insulating region can be obtained by strong-coupling perturbation expansion. 
The strong-coupling expansion becomes valid in the region $U \gtrsim W$ with the bandwidth $W$ (in the square lattice $W =8t$)~\cite{Otsuki_2019}. \Rb and \Br ($A$ = \cation) with $U/t = $9.522 and 10.637, respectively~\cite{Hirayama_2020}, satisfy the condition $U > W$. 

In the strong-coupling expansion, the superexchange interaction $J_{\rm s}$ (with $t^4$-order correction term) and cyclic ring-exchange interaction $J_{\rm c}$ are given by $J_{\rm s} = 4 t^2 / U - 24 t^4 / U^3$ and $J_{\rm c} = 80 t^4 / U^3$, respectively~\cite{Takahashi_1977,MacDonald_1988,Delannoy_2009}. 
If we effectively take into account the effect of the ring-exchange interaction in the nearest-neighbor interaction $J$, the $J$ value becomes 
\begin{eqnarray}
J = J_{\rm s} - 2J_{\rm c} S^2 = \frac{4 t^2} {U} -  \frac{64 t^4 }{U^3}
\end{eqnarray}
with $S=1/2$. 

\subsection{Energy mapping method} 
\label{sec_method_LDA+U}

Within the LDA+$U$~\cite{Anisimov_1991,Anisimov_1993,Liechtenstein_1995,Cococcioni_2012}, we perform the magnetic calculations. Here, $U$ is introduced into the Ni 3$d$ orbital subspace. 
We employ $2\times2\times1$ supercell consisting of four conventional cells. 
We simulate two different magnetic solutions: one is N\'eel type [$(\pi/a,\pi/a,0)$ antiferromagnetic order] and the other is stripe type [$(\pi/a,0,0)$ antiferromagnetic order].
We calculate the energy difference $\Delta E$ between the two antiferromagnetic solutions. 
When we assume the two-dimensional classical spin-1/2 Heisenberg model up to next-nearest-neighbor magnetic interaction $J'$, $\Delta E$ per formula unit is given by $\Delta E = J/2 - J' \simeq J/2$.  
We estimate $J$ with this equation. 

\subsection{Local force approach}
\label{sec_method_Liechtenstein}

Based on the N\'eel-type solutions of the LDA+$U$ calculations, we estimate $J$ and $J'$ using the local force theorem~\cite{Lichtenstein_2013}. 
The local force approach estimates the magnetic interactions from the small energy change induced by the infinitesimal spin-rotation from the magnetic solutions (N\'eel-type in the present case).
We employ the so-called Lichtenstein formula, which is recently developed in the low-energy Hamiltonian with the Wannier orbitals~\cite{Korotin_2015,Nomoto_2020_1,Nomoto_2020_2}, given by,
\begin{align}
    (-1)^PJ_{ij}= 4T\sum_{\omega_n}{\rm Tr}[G_{ij}(\omega_n) M_j G_{ji}(\omega_n) M_i],\label{eq:licht}
\end{align}
where $\omega_n=  (2n+1) \pi T$ denotes the Matsubara frequency. Here, we set $P=0$ (1) when the spins at $i$ and $j$ sites are aligned parallel (anti-parallel) to each other. 
The Green's function $G_{ij}$ is defined by $G_{ij}^{-1}(i\omega_n)=(i\omega_n+\mu)\delta_{ij}-{\mathcal H}^0_{ij}$, 
where ${\mathcal H}^0_{ij}$ is the hopping matrix of the Wannier tight-binding model, and $\mu$ is the chemical potential.
Note that ${\mathcal H}^0_{ij}$ is a $N_{{\rm orb}_i}\times N_{{\rm orb}_j}$ matrix, where $N_{{\rm orb}_i}$ is the number of Wannier orbitals at $i$-site including the spin index. 
In the case of collinear magnets, one may write ${\mathcal H}^0_{ii}$ as ${\mathcal H}^0_{ii}=  \varepsilon_{i}\otimes\sigma_0 + m_i\otimes\sigma_z$. 
Then, $M_i$ is defined by $M_i=m_i \otimes \sigma_x$ and proportional to the exchange splitting at $i$-site $m_i$. 
Here, we have neglected the spin-dependent hopping term of $M_i$ (see Ref.~\onlinecite{Nomoto_2020_1} for details~\footnote{Note that the $J_{ij}$ value in this paper is defined to be eight times as large as that in Ref. ~\onlinecite{Nomoto_2020_1}}).

\subsection{Comparison among the three methods}
\label{sec_method_comparison}
The strong-coupling expansion gives local (in real-space) $J$, the energy mapping method sees the energy difference between the global and local minima of the magnetic solutions, and the local force method sees the low-energy excitations around the global minimum. These $J$ are complementary to each other, and hence we employ all the three methods.
When the Coulomb repulsion is much larger than the bandwidth and the mapping to the Heisenberg model becomes valid, these three methods see the same $J$. 
As we will show in Sec.~\ref{sec_results}, the three results agree reasonably well as expected from a Mott insulating behavior of the proposed $d^9$ nickelates.

\subsection{Calculation conditions}
\label{sec_method_condition}

The DFT band structure calculations are performed using \QE~\cite{QE-2017}.    
We employ Perdew-Burke-Ernzerhof (PBE)~\cite{Perdew_1996} norm-conserving pseudopotentials downloaded from PseudoDojo~\cite{Setten_2018} [the pseudopotentials are based on ONCVPSP (Optimized Norm-Conserving Vanderbilt PSeudopotential)~\cite{Hamann_2013}].

The energy comparison between the N\'eel- and stripe-type antiferromagnetic solutions is performed using $9\times 9 \times 7$ and $9\times 9 \times 3$ {\bf k}-mesh for \Rb and \Br ($A$ = \cation), respectively. 
We treat \cation by the virtual crystal approximation.
The energy cutoff is set to be 100 Ry for the Kohn-Sham wave functions, and 400 Ry for the electron charge density.

For the estimate of $J$ based on the local force approach, we first construct the maximally localized Wannier functions~\cite{Marzari_1997,Souza_2001} for the  N\'eel-type antiferromagnetic band structure using RESPACK~\cite{Nakamura_arXiv,RESPACK_URL}. 
For \Rb, we use $5\times 5 \times 5$ {\bf k}-mesh for the construction of Wannier orbitals. 
We put Ni $d$, O $p$, Ca $d$, and interstitial-$s$ (located at the interstitial positions surrounded by Ni$^{+}$, Ca$^{2+}$, and Rb$^{+}$ cations) projections.
The interstitial orbitals are stabilized because they feel attractions from the surrounded cations~\cite{Nomura_2019}.  
Then, we obtain 104 orbital (per spin) tight-binding Hamiltonian. 
For \Br ($A$ = \cation), we employ $5\times 5 \times 3$ {\bf k}-mesh for constructing Wannier orbitals. 
We derive 232 orbital (per spin) tight-binding Hamiltonian using the projections of Ni $d$, O $p$, Br $p$, $A$ $d$, and interstitial-$s$ (located at the interstitial positions surrounded by Ni$^{+}$, $A^{2.5+}$, and Br$^{-}$ ions) orbitals.

In the calculation of Eq.~\eqref{eq:licht},
we employ $16\times16\times16$ $\bf k$-mesh and set the inverse temperature $\beta= 200$~eV$^{-1}$ for both cases. 
We have confirmed that the difference of $J_{ij}$ values at $\beta= 200$ and 400~eV$^{-1}$ is less than 1 \%.
We use the intermediate representation basis for the Matsubara frequency summation~\cite{Shinaoka_2017,Chikano_2019,Li_Shinaoka_2020}, and set the cutoff parameter $\Lambda=10^5$, which is sufficiently larger than $W\beta$ where $W$ is the band width.

\begin{figure*}[tb]
\vspace{0.0cm}
\begin{center}
\includegraphics[width=0.99\textwidth]{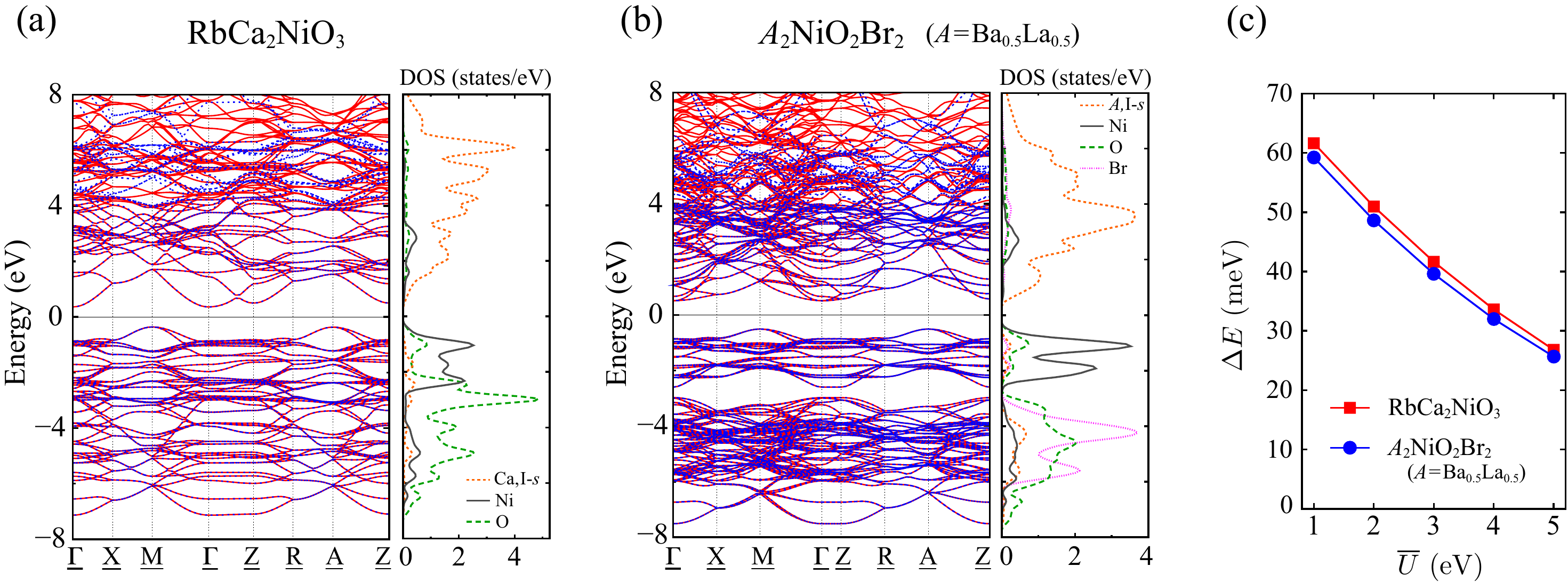}
\caption{
N\'eel-type antiferromagnetic band structure (red curves) and orbital-resolved density of states (per formula unit, per spin) for (a) \Rb and (b) \Br ($A$ = \cation), calculated with LDA+$U$ method ($\overline{U} = 3$ eV). 
The blue dotted curves are band dispersion calculated from the Wannier tight-binding Hamiltonian.
The symbols for the high-symmetry ${\bf k}$ points with the underlines are defined based on $2\times2\times1$ supercell consisting of four conventional cells. 
The origin of the energy axis is set to be the middle of the gap. 
The orbital-resolved density of states is calculated from the Wannier tight-binding Hamiltonian.
``I-$s$" stands for the interstitial-$s$ orbitals
(see Sec.~\ref{sec_method_condition} for the details of the projections used in the Wannier construction). 
(c) The energy difference $\Delta E$ per formula unit between N\'eel- and stripe-type antiferromagnetic solutions. 
The N\'eel-type solutions always show lower energy. 
}
\label{Fig_LDA+U}
\end{center}
\end{figure*}

\section{\mbox{\boldmath$J$} in $d^9$ nickelates}
\label{sec_results}

\begin{figure*}[tb]
\vspace{0.0cm}
\begin{center}
\includegraphics[width=0.99\textwidth]{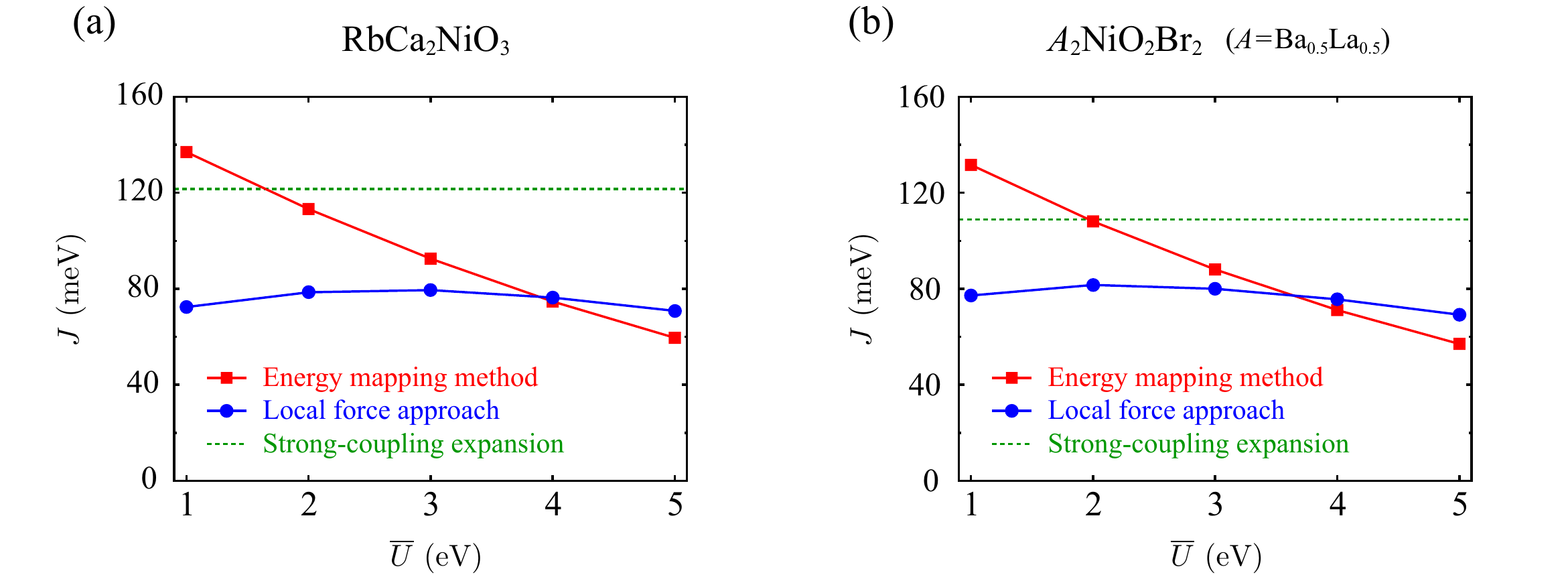}
\caption{
Estimated exchange coupling $J$ for (a) \Rb and (b) \Br ($A$ = \cation). 
$\overline{U}$ is the Hubbard interaction in the LDA+$U$ calculation (the Coulomb repulsion between the Ni 3$d$ orbitals), which we distinguish from the Hubbard $U$ in the single-orbital Hubbard model used in the strong-coupling expansion (the Coulomb repulsion between the Wannier orbitals made from the Ni 3$d_{x^2-y^2}$ orbital with O $2p$ tails). See text for details. 
}
\label{Fig_J}
\end{center}
\end{figure*}

In the previous study~\cite{Hirayama_2020}, the effective single-orbital Hamiltonians for \Rb and \Br ($A$ = \cation) are constructed using maximally-localized Wannier functions~\cite{Marzari_1997,Souza_2001} and cRPA~\cite{Aryasetiawan_2004}.  
The derived nearest-neighbor hopping and Hubbard parameters are 
$t = -0.352$ eV, $U = 3.347$ eV for \Rb, 
and 
$t = -0.337$ eV, $U = 3.586$ eV for \Br ($A$ = \cation). 
Then, the strong-coupling expansion described in Sec.~\ref{sec_method_superexchange} gives 
$J = 122$ meV and $J=109$ meV for \Rb and \Br ($A$ = \cation), respectively 
(see Appendix~\ref{Appendix_dp} for the estimate from three-orbital $d$-$p$ model).

Figures~\ref{Fig_LDA+U}(a) and \ref{Fig_LDA+U}(b) show the band structure calculated by the LDA+$U$ method for the
N\'eel-type antiferromagnetic state.
While the Hubbard $U$ in the single-orbital Hubbard model is the Coulomb repulsion between the Wannier orbitals made from the Ni 3$d_{x^2-y^2}$ orbital with O $2p$ tails,
the $U$ interaction in the LDA+$U$ calculation is the Coulomb repulsion between the Ni 3$d$ orbitals. 
To make the difference clearer, we call $U$ in the LDA+$U$ calculation $\overline{U}$. 
In Figs.~\ref{Fig_LDA+U}(a) and \ref{Fig_LDA+U}(b), we have used $\overline{U}$ = 3 eV.

In contrast to the case of the LDA+$U$ calculation for NdNiO$_2$, where the system stays metal even in the presence of antiferromagnetic order~\cite{Botana_2020,HuZhang_2020,Z_Liu_2020}, both systems become insulating. 
The top of the valence band has mainly Ni $3d$ character, in agreement with the classification into the Mott-Hubbard type insulator. 
We see that both systems are insulating even at smaller $\overline{U}$ (= 1 eV). 
For all the $\overline{U}$ region we studied (1-5 eV), there exists well defined spin-1/2 Ni spin moment. 
The results suggest that, if these $d^9$ nickelates are synthesized, they become antiferromagnetic Mott insulator as in the cuprates. 

Figure~\ref{Fig_LDA+U}(c) shows the energy difference $\Delta E$ per formula unit between the N\'eel- and stripe-type antiferromagnetic solutions. 
$\Delta E$ decreases as $\overline{U}$ increases, which is a natural behavior given that $\Delta E$ is governed by $J$ and the origin of $J$ is the superexchange interaction. 

In Figs.~\ref{Fig_LDA+U}(a) and \ref{Fig_LDA+U}(b), the band dispersions obtained by the Wannier tight-binding Hamiltonian, which are used in the local force approach, are also shown. 
The Wannier bands well reproduce the LDA+$U$ magnetic band dispersions.

From $\Delta E$ in Fig.~\ref{Fig_LDA+U}(c), we perform the order estimate of $J$ by the energy mapping method with assuming $J'/J = 0.05$ (Sec.~\ref{sec_method_LDA+U}) \footnote{We do not pay special attention to the precise value of the ratio $J'/J$ because we are only interested in the order estimate of $J$.}. 
Then $J$ is given by $J = \Delta E / 0.45 $.
We also estimate $J$ using the local force approach (Sec.~\ref{sec_method_Liechtenstein}). 

These results on top of the $J$ value estimated by the strong-coupling expansion (see above) are summarized in Figs.~\ref{Fig_J}(a) and \ref{Fig_J}(b) for \Rb and \Br ($A$ = \cation), respectively. 
The $J$ value in the energy mapping method changes from about 140 meV ($\overline{U}=1$ eV) to 60 meV ($\overline{U}=5$ eV). 
The local force approach gives $J \simeq 70$-80 meV. 
These estimates give the same order of $J$ as the strong-coupling expansion results [$J = 122$ meV and $J=109$ meV for \Rb and \Br ($A$ = \cation), respectively].

Although the energy mapping method and local force approach are based on the same LDA+$U$ calculations, 
we see that there is a discrepancy between the two results at small $\overline{U}$ values (although the difference is no more than two times).
It should be noted that the former method sees the global change of the energy between the completely different magnetic patterns, 
whereas the latter approach only sees the local landscape around the N\'eel-type solutions, 
as described in Sec.~\ref{sec_method_comparison}. 
For larger $\overline{U}$, the agreement between these two results becomes better as is expected: The system can be mapped to the classical spin model with a constant $J$ regardless of the assumed magnetic structure in the local force approach.

Overall, all the three estimates of $J$ lie within 60-140 meV, and we conclude that the $d^9$ nickelates have a sizable $J$ of the order of 100 meV. 
The agreement in the order estimate of $J$ among three independent methods shows that 
\Rb and \Br ($A$ = \cation) are indeed Mott insulators with the effective model being the Heisenberg model, and the magnetic exchange coupling $J$ is governed by the superexchange interaction [if the materials were, for example, weakly correlated, the three methods would not agree well (see Sec.~\ref{sec_method_comparison})].

Finally, we compare the $J$ value with that of the cuprates. 
In the cuprates, the magnitude of $J$ is intensively studied by Raman spectroscopy in the early stage~\cite{Lyons_1988_1,Lyons_1988_2,Sugai_1988}. 
The $J$ value for La$_2$CuO$_4$ is estimated to be about 130 meV~\cite{Singh_1989}.
Systematic investigations have shown that the material dependence of $J$ in the cuprates family is weak~\cite{Sulewski_1990,Tokura_1990_raman}.
The numerical study on the $d$-$p$ model has also derived $J$ as large as about 130 meV~\cite{Hybertsen_1990}. 
Compared to the $J$ value of 130 meV for the cuprates, our estimate based on the $d$-$p$ model giving 90-100 meV (see Appendix~\ref{Appendix_dp}) is small, which is consistent with the fact that $\Delta_{dp}$ is larger in the nickelates. 
However, we note that the $J$ value of about 100 meV is still significantly large, and the $d^9$ nickelates would serve as interesting cuprate-analog materials.

\section{Summary}
\label{sec_summary}

One of the remarkable features of the high $T_{\rm c}$ cuprates is the large exchange coupling $J$, whose size is as large as 130 meV~\cite{Lee_Nagaosa_Wen_2006}. In the present study, we have evaluated the size of $J$ for $d^9$ nickelates from first principles.
While the cuprates having small $\Delta_{dp}$ belong to the charge-transfer type in the Zaanen-Sawatzky-Allen diagram~\cite{Zaanen_1985}, nickelates with larger $\Delta_{dp}$ belong to the Mott-Hubbard type. 
To answer how large $J$ can be expected in the Mott-Hubbard insulating $d^9$ nickelates, we studied \Rb and \Br ($A$ = \cation), 
which were recently proposed theoretically and shown to be free from the self-doping in Ref.~\onlinecite{Hirayama_2020}. 
By means of the strong-coupling expansion, energy mapping method, and local force approach, we found that $J$ for these nickelates is as large as 100 meV, which is not far smaller than that of the cuprates. 
This result suggests that the $d^9$ nickelates and cuprates share a notable common feature in the Mott insulating phase, although the former and latter belong to the Mott-Hubbard and charge-transfer regime, respectively.


Finally, we note that the proposed $d^9$ nickelates might give rare examples of realizing the square-lattice Hubbard model with sizeable $J$ in real materials. 
Recent numerical studies show that the phase diagram of the doped Hubbard model is under severe competition between the stripe state with charge/spin modulation and $d$-wave superconductivity~\cite{Zheng_2017,Darmawan_2018,Ohgoe_2020,HC_Jiang_2019}. 
Therefore, once synthesized, the $d^9$ nickelates will serve as a valuable test-bed system to understand the superconductivity in the Hubbard-like model. 
They are also an important reference to understand the superconducting mechanism in the cuprates, because they would tell us whether the charge-transfer nature in the cuprates is essential in the high-$T_{\rm c}$ superconductivity or not.

\begin{acknowledgments}
We acknowledge the financial support by JSPS KAKENHI  
Grant No. 16H06345 (YN, MH, and RA), 17K14336 (YN), 18H01158 (YN), 19K14654 (TN), 19H05825 (RA), 20K14390 (MH), and 20K14423 (YN). 
This work was supported by MEXT as ``Program for Promoting Researches on the Supercomputer Fugaku" (Basic Science for Emergence and Functionality in Quantum Matter).
A part of the calculations was performed at Supercomputer Center, Institute for Solid State Physics, University of Tokyo.
\end{acknowledgments}

\appendix

\section{Exchange coupling $J$ from $d$-$p$ model}
\label{Appendix_dp}

In the main text, we estimate $J$ by the strong-coupling expansion starting from the single-band Hubbard model. 
Here, we show that the $J$ value is also on the order of 100 meV even when we perform the strong-coupling expansion based on the so-called $d$-$p$ model consisting of Ni 3\xx and two O $2p$ orbitals. 
In the strong-coupling expansion of the $d$-$p$ model for the filling of one hole per unit cell, $J$ is given by 
\begin{eqnarray}
J= \frac{4t_{dp}^4}{\Delta_{dp}^2 U_{dd} } + \frac{4 t_{dp}^4}{\Delta_{dp}^2 (\Delta_{dp} + U_{pp}/2 ) }, 
\end{eqnarray}
where $t_{dp}$ is the hopping between Ni 3\xx and O $2p$ orbitals, $U_{dd}$ and $U_{pp}$ are the onsite Coulomb repulsion for Ni 3\xx and O $2p$ orbitals, respectively.

Using the RESPACK~\cite{Nakamura_arXiv,RESPACK_URL} based on the cRPA method~\cite{Aryasetiawan_2004,Sasioglu_2011} combined with the maximally-localized Wannier functions~\cite{Marzari_1997,Souza_2001}, we constructed three-orbital $d$-$p$ model from first principles. 
We consider the double counting effect in the Hartree term and $\Delta_{dp}$ is given by  $\Delta_{dp} = \Delta_{dp}^{\rm DFT} + U_{dd} \ \! \underline{n}_d^{\rm DFT}/2 - U_{pp} \ \! \underline{n}_p^{\rm DFT}/2$, where the superscript DFT stands for the DFT value, and $\underline{n}$ is the hole occupation.

For \Rb, 
we obtain $ | t_{dp} | = 1.23$ eV, $\Delta_{dp} = 5.46$ eV ($\Delta_{dp}^{\rm DFT} = 4.11$ eV), $U_{dd} = 4.83$ eV, and $U_{pp} = 4.62$ eV.
Then the $J$ value is estimated as $J=103$ meV.

For \Br ($A$ = \cation), 
we get $ | t_{dp} | = 1.24$ eV, $\Delta_{dp} = 5.86$ eV ($\Delta_{dp}^{\rm DFT} = 4.37$ eV), $U_{dd} = 5.05$ eV, and $U_{pp} = 4.57$ eV.
The resulting $J$ value is $J= 88$ meV. 



\bibliography{apssamp}

\end{document}